\documentclass[%
 aip,
 jap,%
 amsmath,amssymb,
 reprint,%
]{revtex4-1}

\usepackage{graphicx}
\usepackage{dcolumn}
\usepackage{bm}

\begin{document}

\title[Sample title]{Anomalous Hall effect in NiPt thin films}

\author{T. Golod}
 \email{taras.golod@fysik.su.se}
\author{A. Rydh}
\author{V. M. Krasnov}%
\affiliation{Department of Physics, Stockholm University, AlbaNova University Center, SE~--~106 91 Stockholm, Sweden}%

\date{\today}

\begin{abstract}
We study Hall effect in sputtered Ni$_{x}$Pt$_{1-x}$ thin films
with different Ni concentrations. Temperature, magnetic field and
angular dependencies are analyzed and the phase diagram of NiPt
thin films is obtained. It is found that films with sub-critical
Ni concentration exhibit cluster-glass behavior at low
temperatures with a perpendicular magnetic anisotropy below the
freezing temperature. Films with over-critical Ni concentration
are ferromagnetic with parallel anisotropy. At the critical
concentration the state of the film is strongly frustrated. Such
films demonstrate canted magnetization with the easy axis rotating
as a function of temperature. The magnetism appears via
consecutive paramagnetic - cluster glass - ferromagnetic
transitions, rather than a single second-order phase transition.
But most remarkably, the extraordinary Hall effect changes sign at
the critical concentration. We suggest that this is associated
with a reconstruction of the electronic structure of the alloy at
the normal metal - ferromagnet quantum phase transition.

\end{abstract}

\maketitle

\section{Introduction}
NiPt is an interesting alloy both from fundamental and
technological points of view, due to non-trivial magnetic and
catalytic properties\cite{catalytic1,catalytic2}. Pt is
characterized by the strong spin-orbit interaction resulting in
the appearance of a large spin-Hall effect \cite{Vila,Guo}. When
mixed with magnetic elements, like Ni or Fe, Pt provides a
non-trivial host matrix leading to a variety of physical
properties of binary alloys. Detailed understanding of those
properties remains a serious theoretical challenge. It is further
complicated by the existence of several ordered phases, showing
different properties compared to disordered alloys
\cite{Dahmani,Ruban,Olovsson,Kumar(ordered)}. Pt and Pd are the
two known elements that can form solid solutions with Ni at
arbitrary proportions \cite{Besnus,Dahmani} (all three belong to
the same group-10 of the periodic table of elements). However,
while NiPd becomes ferromagnetic at very low Ni concentration, the
onset of zero temperature ferromagnetism in bulk
Ni$_{x}$Pt$_{1-x}$ occurs at fairly large concentration $x_c\simeq
40$ at.$\%$ of Ni\cite{Besnus, Kumar(ordered),Alberts}, which
makes it easier to control composition, material and magnetic
properties. At the critical concentration, the transition between
the normal metal and the ferromagnetic states occurs via a quantum
phase transition \cite{Varma,Vojta}, i.e., a phase transition at
$T=0$, not driven by thermal fluctuation.

Magnetic moment distribution in bulk NiPt alloys was studied quite
exhaustively by high field susceptibility \cite{Beille},
magnetization \cite{Alberts} and neutron scattering \cite{neutron}
experiments. It was observed that NiPt alloys indeed are spatially
uniform, in contrast to other binary alloys of Ni and Fe, which
are prone to segregation.\cite{Beille} The bulk NiPt alloy can
exist in two phases: a chemically disordered face centered cubic
(fcc), with Pt and Ni atoms randomly distributed over the crystal
lattice; and a chemically ordered state, either with face centered
tetragonal (fct), or fcc structure \cite{Dahmani}. Concentration
dependencies of the magnetic moment and the Curie temperature are
different for ordered and disordered alloys, especially at low Ni
concentrations. \cite{Alberts, Kumar(ordered), Kumar(disordered)}

In this work we study magnetic properties of sputtered
Ni$_{x}$Pt$_{1-x}$ thin films with different Ni concentration. The
composition of the films is characterized by the energy-dispersive
X-ray spectroscopy. Anomalous Hall effect is employed for analysis
of magnetic properties of the films. Temperature, magnetic field
and angular dependencies of the Hall resistance are analyzed and
the magnetic phase diagram of Ni$_{x}$Pt$_{1-x}$ thin films is
obtained. It is found that films with low, sub-critical, Ni
concentration show cluster-glass (CG) behavior at low temperatures
and exhibit perpendicular magnetic anisotropy below the freezing
temperature. Films with over-critical Ni concentration are
ferromagnetic (FM) with parallel anisotropy. At the critical
concentration the state of the film is strongly frustrated:
magnetization is canted and is rotating with temperature.
Magnetism appears via a percolative paramagnet - cluster glass -
ferromagnet transitions, rather than a single second order phase
transition. But most remarkably, the extraordinary Hall
coefficient changes sign from electron-like to hole-like at the
critical concentration, while the ordinary Hall coefficient
remains always electron-like. This indicates the ``intrinsic"
nature of the anomalous Hall effect in this case. We suggest that
this phenomenon is a consequence of the quantum phase transition,
and is associated with a reconstruction of the electronic
structure upon transition to the spin-polarized FM state.

In recent years ferromagnetic thin films have attracted
significant attention, due to a rapid development of novel
spintronic applications \cite{Zutic}. In particular, diluted
ferromagnets are favorable for fabrication of hybrid
superconductor/ferromagnet quantum devices, which may benefit both
from spin polarization in FM and macroscopic phase coherence of
superconductors \cite{Demler,
Buzdin,Ryazanov,Aprili,Gu,Krasnov,Frolov,Golod,Feofanov}. The
uniformity of the FM alloy can become critical when designing
small, mono-domain spin-valve type structures. Here NiPt may be
the material of choice, among binary alloys, because of its
inherent intrinsic homogeneity.

Magnetic properties of thin films can be quite different from
those of bulk materials. For example, magnetic moment of
transition metals is predicted to be higher at the surface than in
the bulk\cite{Freeman}. Magnetic properties of thin films may also
depend on the mechanical stress, induced by the substrate. The
magnetic moment can increase or decrease depending on whether the
crystallographic unit cell of the film dilates or contracts, as a
result of the lattice mismatch between the film and the substrate.
The Curie temperature of thin films can also be different compared
to the bulk material\cite{magnetism}. Ferromagnetic thin films
exhibit interesting anisotropy properties. The magnetic moment
usually tends to orient itself in-plane, to minimize the
magnetostatic energy. However, it was shown that thin films of
rare-earth transition-metal alloys may have a perpendicular,
out-of-plane anisotropy \cite{peprpendicular_anisotropy1}. A
significant structural anisotropy is needed for overcoming the
large shape anisotropy, inherent to thin films, in order to
facilitate the out-of-plane easy axis direction of magnetization.
Reorientation of magnetization from the in-plane to the out-of-plane
as a function of film thickness or temperature was observed for a
large number of thin films
\cite{pa3,*pa4,pa5,pa6,pa7,pa8,pa9,pa10}. Films with perpendicular
anisotropy can be used for high-density magnetic recording.

Magnetic thin films may exhibit superparamagnetic-type
of cluster-glass behavior \cite{superparamegnetism1,
superparamegnetism}, which is not common for bulk materials. In
the CG state the system behaves not like one single domain
particle with all its moments aligned in one direction, but rather
splits into multiple spin-clusters. The effective
magnetic moment of the cluster can be large
\cite{superparamegnetism} $\sim 1000$ Bohr magnetons $\mu_B$. At
high temperatures the direction of magnetization of each cluster
fluctuates and eventually freezes at low $T$. Below freezing
temperature the system can exhibit a spontaneous magnetization and
coercivity.

The anomalous Hall effect \cite{Hurd,Nagaosa,eheNifilms} is the
characteristic property of ferromagnetic materials caused by
spin-orbit interactions. It may have both extrinsic and intrinsic
contributions, arising respectively from spin-dependent impurity
scattering, or finite effective magnetic flux, associated with the
Berry phase of itinerant charge carriers with different spin
polarization \cite{EHE,Nagaosa}. Hall voltage of thin films is
described by\cite{Hurd}

\begin{equation}
V_{H}=(\rho_{H}I)/d=(R_{0}H+R_{1}M)I/d, \label{eq1}
\end{equation}
where $V_{H}$ is the Hall voltage, $\rho_{H}$ is the Hall
resistivity, $H$ is the magnetic field intensity, $M$ is the
magnetization, $I$ is the applied current, $d$ is the film
thickness, $R_{0}$ is the ordinary Hall effect (OHE) coefficient and $R_{1}$ is the extraordinary Hall effect (EHE) coefficient. Magnetic alloys, including Pt-based thin films, have been reported to show anomalously large EHE \cite{ehe1,ehe2,ehe3,PtEHE}, which
is up to two orders of magnitude larger than for magnetic elements such
as Fe, Co and Ni \cite{Watanabe}. The EHE provides a very simple
way of studying magnetic properties of thin films, compared to
other measurement techniques such as vibrating sample
magnetometer\cite{VSM}, superconducting quantum interference
device magnetometer\cite{SQUID}, optical
\cite{optical_magnetometer} and Hall probe magnetometer
\cite{EHE}. The reciprocal dependence of the measured Hall voltage
on the film thickness, see Eq.(\ref{eq1}), makes this technique
preferential for analysis of thin films. The EHE allows us to
study magnetic properties at all temperatures and magnetic
fields.\cite{EHE_films}

\begin{figure}[t]
\includegraphics[width=18pc]{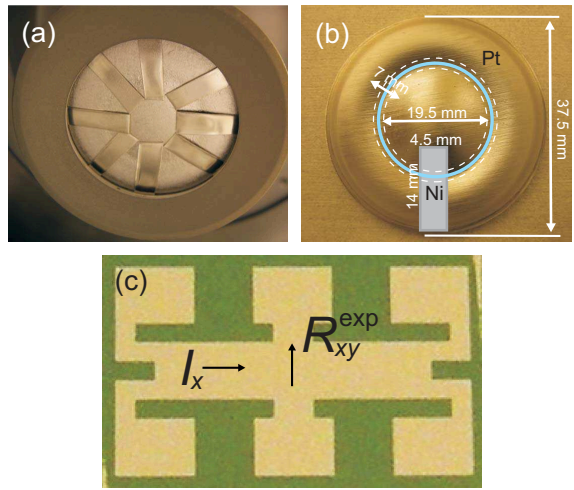}
\caption{\label{f01} (Color online) (a) The sputtering Pt target
with attached Ni segments. (b) The erosion track at the 1.5 inch
Pt target. The position of rectangular Ni segment with the size
$4.5\times 14$ mm is shown schematically. Each Ni segment covers
$\sim 7.4 \%$ of the effective Pt target area. (c) The NiPt Hall
probe bridge. The current is applied along the sample and
resistance is measured simultaneously in longitudinal and
transverse direction.}
\end{figure}

\section{Experimental}

\subsection{Thin film fabrication}

Ni$_{x}$Pt$_{1-x}$ thin films were deposited by DC magnetron
sputtering from 1.5 inch target on oxidized Si wafer at a power of
0.05 kW, base pressure $\sim 10^{-6}$ mbar and processing Ar
pressure $6.7\cdot10^{-3}$ mbar. The deposition time was about 5
minutes with deposition rate of 1.67 {\AA}/sec. The thickness of
all studied films is $\sim 50-60$ nm, as determined by surface
profilometer measurements.

Due to the large cost of Pt, preparation of separate NiPt sputter
targets for different Ni concentrations is impractical. Therefore,
deposition targets were made from a single pure Pt target, on top
of which a selected number of rectangular shaped Ni segments were
symmetrically attached, similar to that in Ref.~\onlinecite{Sato}.
The number of segments was adjusted to control the composition of
the studied Ni$_{x}$Pt$_{1-x}$ thin films. Figure \ref{f01}(a)
shows the target with eight attached Ni segments. Each rectangle
covers $\sim7.4~\%$ of the effective sputtering area, confined
within the erosion track with the diameter of 19.5 mm and the
width 7 mm, as shown in Fig. \ref{f01}(b). The deposited
Ni$_{x}$Pt$_{1-x}$ films were patterned using photolithography and
Ar ion milling to form a bridge with both Hall and longitudinal
contacts, as shown in Fig. \ref{f01}(c).

\begin{figure}[t]
\includegraphics[width=18pc]{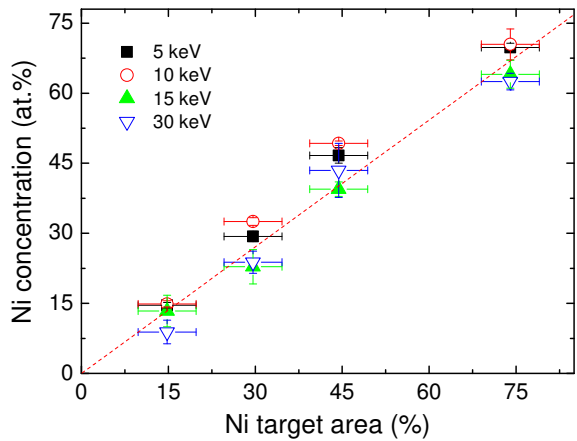}
\caption{\label{f02} (Color online) EDS measured Ni concentration
of the NiPt thin films on SiO$_{2}$ substrates as a function of
the relative Ni target area for different electron beam energies.
The dashed line represents a linear fit for different electron
beam energies.}
\end{figure}

\subsection{Analysis of thin films composition by energy dispersive X-ray spectroscopy}

To determine the actual Ni concentration in our NiPt thin films,
we used the energy dispersive X-ray spectroscopy (EDS). Electron
probe microanalysis showed uniform composition of the film with
the spatial resolution $\sim 1 \mu m$. Both K and L shell X-ray
series were analyzed \cite{Golod1}. Analysis of thin film
composition by EDS is non-trivial because incident electrons can
shoot through the film and cause a parasitic X-ray signal from the
substrate \cite{EDS1, EDS2}. This may cause an error due to
secondary fluorescence of the film material after reabsorption of
X-rays produced in the substrate. Furthermore, the film/substrate
interface leads to backscattering of electrons into the film,
which causes additional X-ray emission and overestimation of
concentration of elements in the film. Depending on the electron
beam energy, the interaction volume can be confined either in the
film or in the substrate, resulting in different measured X-ray
signals. Thus, the conventional quantitative EDS correction, used
for bulk specimens, can give errors when applied to thin films.

To check the influence of the interaction volume on the EDS
signal, we varied electron beams energies. Figure \ref{f02} shows
obtained Ni concentrations as a function of the relative area of
Ni in the combined Pt/Ni target for electron beam energies of 5,
10, 15 and 30 keV. The effective target area was calculated taking
into account the actual profile of the erosion track on the
target, shown in Fig. \ref{f01}(b). It is seen that the measured
Ni concentration is correlated with the relative Ni target area.
However, EDS measurements at different E-beam energies provide
slightly different estimations of Ni concentration. The largest
value of Ni concentration is obtained at 10 keV energy for films
at SiO$_{2}$ substrates.

For lower electron beam energy of 5 keV the interaction volume is
mostly confined in the film, consequently the number of
backscattered electrons at the film/substrate interface is small.
At high electron beam energies 15 and 30 keV, the number of
backscattered electrons from the interface is also small because
the peach-form interaction volume is confined mostly in the
substrate and the probability for high energy incident electrons
to be reflected at the film/substrate interface is low. However,
at the intermediate electron beam energy of 10 keV, the extremum
of the interaction volume cross-section is close to the
film/substrate interface. The energy of incident electrons in this
case is low enough for having a large probability of
backscattering at the film/substrate interface, simultaneously
backscattered electrons have sufficiently high energy for
excitation of L-series X-rays of Ni, which are used for EDS
analysis at this beam energy. This can lead to a particularly
large overestimation of Ni concentration at 10 keV E-beam energy.

\begin{figure}[t]
\includegraphics[width=18pc]{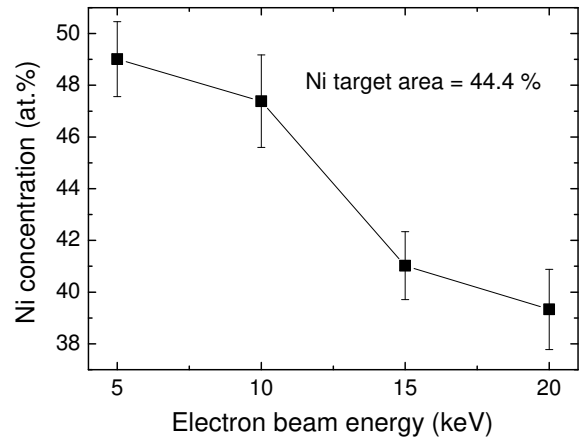}
\caption{\label{f03} EDS measured Ni concentration of NiPt flakes
as a function of electron beam energy. The flakes were scratched
out from a film with 44.4 $\%$ of the relative Ni target area.}
\end{figure}

In order to eliminate the effect of the substrate, we performed
EDS analysis on free standing NiPt flakes, spread on a carbon
substrate. Figure \ref{f03} shows the average value of Ni
concentration for several NiPt flakes, as a function of the
electron beam energy. The film was deposited from the Pt/Ni target
with the relative Ni area of 44.4 $\%$. It is seen that in this
case the estimated Ni concentration decreases monotonously with
increasing the electron beam energy. This can not be attributed to
backscattering from the substrate but rather can be explained by
the error in correction of Ni X-ray absorption in the film
\cite{EDS2} or/and by the error in the correction of the secondary
fluorescence of Ni by Pt. At low electron beam energies, the light
L-series are used for determination of Ni concentrations, while
high energy K-series are used at high electron beam energies. The
absorption is very difficult to correct when working with light
elements. It is also more difficult to correctly compensate the
secondary fluorescence of very light Ni L-series by Pt. Hence, Ni
concentrations are obtained from the linear fit to EDS data at
electron beam energies of 5, 10, 15 and 30 keV but with lower
weight for 10 keV and larger weight for 15 and 30 keV (the dashed
line in Fig. \ref{f02}). In what follows, we will use this fit for
notation of Ni concentration. The accuracy of such determination
is verified below by comparison of the Curie temperature at high
67\% Ni concentration with that for bulk alloys.

\subsection{Measurement setup}

\begin{figure}[t]
\includegraphics[width=18pc]{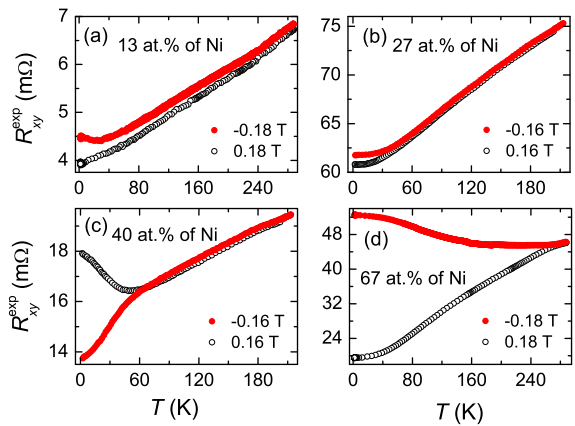}
\caption{\label{f05} (Color online) Temperature dependencies of transverse resistances for the films with Ni concentrations of (a) 13, (b) 27, (c) 40 and (d) 67 at.$\%$ and for different field directions ($\pm0.16$ and $\pm0.18$ T) perpendicular to the
film.}
\end{figure}

Measurements were carried out in a cryogen-free magnet system with
a flowing gas insert, in a temperature range from 1.8 K to 300 K
and fields up to $\pm$10 T. The sample was mounted on a rotatable
sample holder. Resistances were measured both in transverse
($R^{\mathrm{exp}}_{xy}$) and longitudinal directions
($R^{\mathrm{exp}}_{xx}$) simultaneously at constant current bias
in longitudinal direction ($I_{x}$), as shown in Fig.
\ref{f01}(c). The possibility to rotate the platform makes it
possible to apply magnetic field at any angle with respect to the
film surface. All studied films have about the same thickness and
the same Hall-bridge geometry. Therefore, presented data for films
with different Ni concentrations can be compared explicitly
without the need of additional geometrical normalization factors.

Due to the specific geometry of the experiment, see Fig.
\ref{f01}(c), the measured $R^{\mathrm{exp}}_{xy}$ is sensitive only to the
perpendicular component of the magnetic induction in the film. For
the out-of-plane orientation of the easy axis, magnetization in
the perpendicular field will have a hysteresis/coercivity and so
will $R^{\mathrm{exp}}_{xy}$. However, for the in-plane orientation of the easy
axis, the magnetization of the film in perpendicular magnetic
field will have zero coercivity, because in this case only
rotation of magnetization takes place without translational
movement of magnetic domain walls.\cite{film_anisotropy}

\section{Results and discussion}

\begin{figure}[t]
\includegraphics[width=18pc]{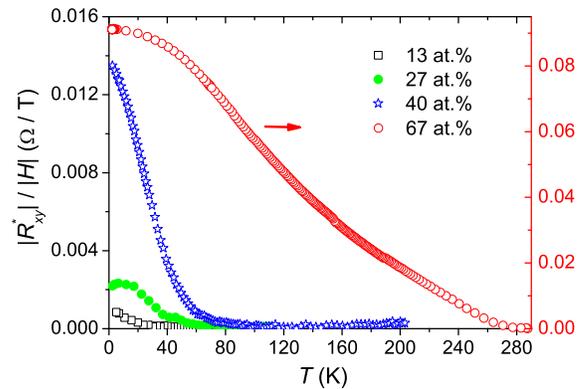}
\caption{\label{f06} (Color online) Absolute values of Hall
resistances, normalized by field, as a function of temperature for
films with different Ni concentrations. The $R_{xy}^*$ is obtained
from the data in Fig. \ref{f05}, using Eq. (\ref{eq2}).}
\end{figure}

Figure \ref{f05} shows temperature dependencies of measured
transverse resistances $R^{\mathrm{exp}}_{xy}$ for films with Ni
concentrations of 13, 27, 40 and 67 at.$\%$ and for two opposite
field directions perpendicular to the film.
$R^{\mathrm{exp}}_{xy}$ contains both longitudinal and Hall
contributions. It is seen that at high $T$ the measured
$R^{\mathrm{exp}}_{xy}$ is independent of the field direction,
indicating that it is dominated by the even-in-field longitudinal
resistance. However, at lower $T$ magnetic correlations become
significant and the corresponding odd-in-field Hall contribution
appears in $R^{\mathrm{exp}}_{xy}$. To extract the pure Hall
contribution, we took the difference of $R^{\mathrm{exp}}_{xy}$
for positive and negative field directions:
\begin{equation}
R^{*}_{xy}=\frac{[R^{\mathrm{exp}}_{xy}(+H)-R^{\mathrm{exp}}_{xy}(-H)]}{2}. \label{eq2}
\end{equation}
This way the contribution from the even in field longitudinal
resistance is cancelled out. Alternatively, we explicitly subtract
the longitudinal resistance $R^{\mathrm{exp}}_{xx}$, measured
simultaneously:
\begin{eqnarray}\label{eq2b}
R_{xy}=R^{\mathrm{exp}}_{xy}-\beta R^{\mathrm{exp}}_{xx},\\
\beta =\frac{R^{\mathrm{exp}}_{xy}(H=0)}{R^{\mathrm{exp}}_{xx}(H=0)}.
\end{eqnarray}
The advantage of such definition is that it provides Hall
resistances for both field directions, while Eq.(\ref{eq2}) gives
$R^{*}_{xy}$ only for the absolute value of the field.

\begin{figure}[t]
\includegraphics[width=18pc]{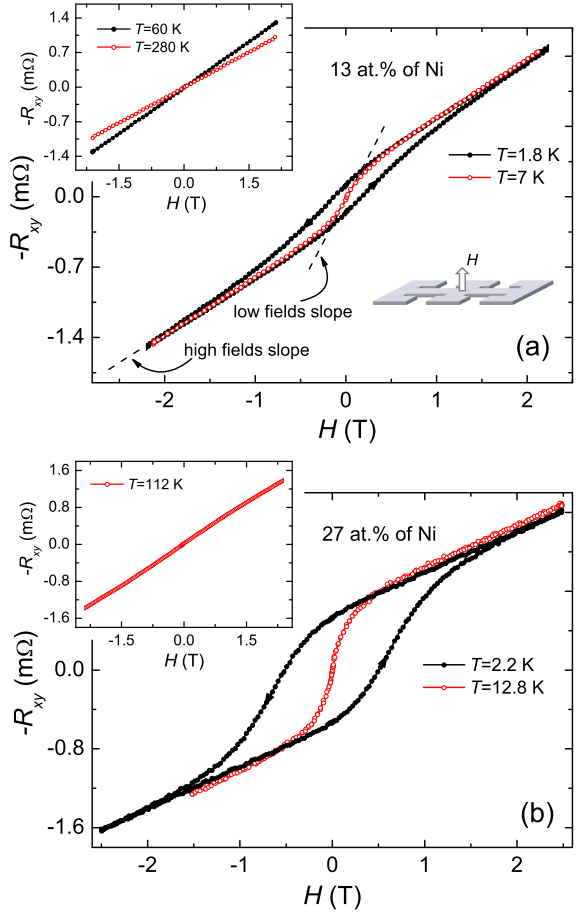}
\caption{\label{f07} (Color online) Magnetic field dependencies of
Hall resistances at low (main panels) and elevated (insets)
temperatures for films with (a) 13 at.$\%$ and (b) 27 at.$\%$ of
Ni. The field is perpendicular to the films. At low $T$ (solid
circles), $R_{xy}(H)$ for both films exhibits hysteresis. At
higher $T$ the films are in the cluster-glass state (open
circles): $R_{xy}(H)$ remain strongly nonlinear, but the
coercivity vanishes at $T \simeq 7$ K for 13 at.$\%$ Ni and $T
\simeq 12.8$ K for 27 at.$\%$ of Ni. With further increase of $T$,
the films become paramagnetic with linear $R_{xy}(H)$, as shown in
insets.}
\end{figure}

Figure \ref{f06} shows $T$ dependencies of absolute values of
$R^{*}_{xy}/H$, for the data from Fig. \ref{f05}. An onset of
$R^{*}_{xy}(T)$ with decreasing temperature is seen for all
studied films. The $|R^{*}_{xy}|/|H|$ has strong $T-$dependence
and increases rapidly with increasing Ni concentration. At $T=2$K,
$|R^{*}_{xy}|/|H|$ differs by more than two orders of magnitude
between films with the largest (67 at.$\%$) and the smallest (13
at.$\%$) Ni concentrations.

\subsection{Films with low Ni concentration}

\begin{figure}[t]
\includegraphics[width=18pc]{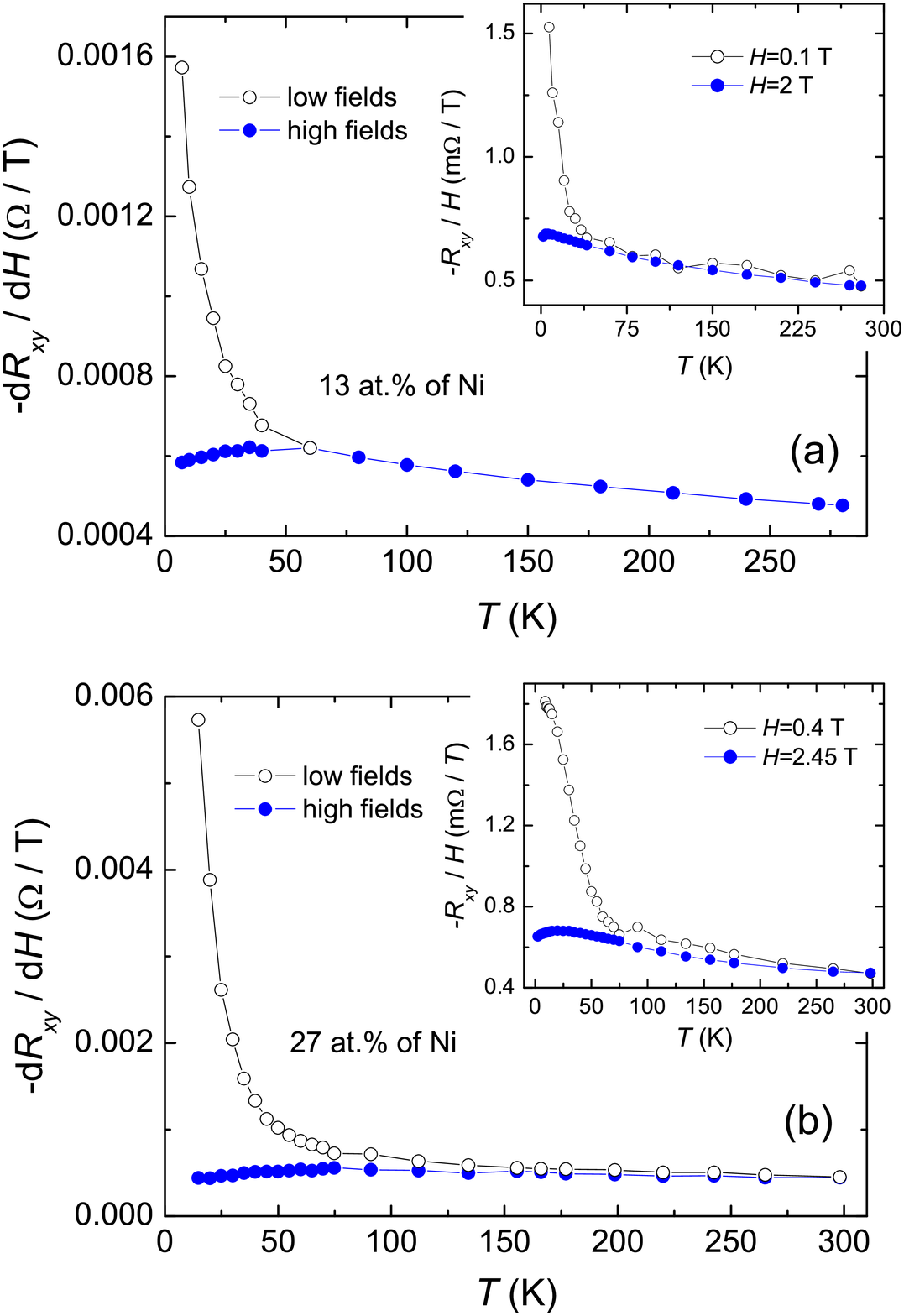}
\caption{\label{f08} (Color online) Temperature dependencies of
$-dR_{xy}/dH$ at $H\simeq 0$ (open) and $H$=2 T (solid symbols)
for films with (a) 13 at.$\%$ and (b) 27 at.$\%$ of Ni. The insets
show $T-$ dependence of normalized Hall resistances at low and
high fields. Existence of the extended cluster-glass region with
nonlinear $R_{xy}(H)/H \neq \mathrm{const}$ is clearly seen.}
\end{figure}

Main panels of Fig. \ref{f07} show $-R_{xy}(H)$ at low $T$ for
films with (a) 13 at.$\%$ and (b) 27 at.$\%$ of Ni. Both films
show a hysteresis at the lowest $T$ (solid circles) with
significant coercivity of 0.16 T and 0.6 T, respectively,
indicating the out-of-plane magnetic anisotropy. It is seen,
however, that $R_{xy}(H)$ continues to grow at large fields. This
is due to a weak magnetism in those strongly diluted, sub-critical
NiPt films. In this case the saturable EHE is of the same order as
the non-saturable OHE contribution.\cite{OHEoverEHE} With
increasing temperature, the coercivity rapidly decreases and
vanishes at $T_f\simeq 7$ K for the film with 13 at.$\%$ of Ni and
$T_f \simeq 12.8$ K for 27 at.$\%$ of Ni. However, $R_{xy}(H)$
remains nonlinear, see curves with open circles in the main panels
of Fig. \ref{f07}, indicating presence of the residual cluster
magnetism in the films. Therefore, $T_f$ represents the CG
freezing temperature.

\begin{figure}[t]
\includegraphics[width=18pc]{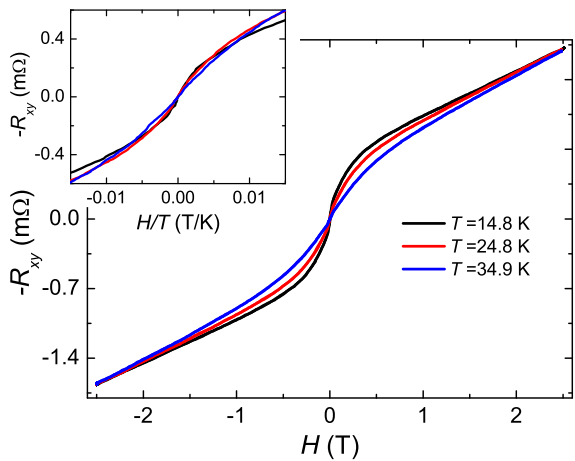}
\caption{\label{f11} (Color online) $-R_{xy}(H)$ dependencies of
the film with 27 at.$\%$ of Ni measured at $T$=14.8, 24.8 and 34.9
K. The inset shows the same $R_{xy}$ as a function of $H/T$ in the
small field range. A good superparamegnetic-type scaling is seen.}
\end{figure}

Figure \ref{f08} shows $-dR_{xy}/dH$ as a function of temperature
for the same films at low fields (open circles) and above the
saturation field at $H$=2 T (solid circles). According to Eq. (1),
the latter represents the OHE. The difference of $dR_{xy}/dH$ at
low and high fields provides an accurate measure of nonlinearity
of the Hall effect. From Fig. \ref{f08} it is seen that at low
temperatures, the zero-field slope has a strong $T-$dependence. It
rapidly decreases with increasing temperature and becomes equal to
the high field slope at $T\simeq 60$ K for the film with 13
at.$\%$ Ni and $T\simeq 112$ K for 27 at.$\%$ Ni. Those
temperatures represent the cluster glass-to-paramagnetic
transition in the films. At higher temperatures $R_{xy}(H)$
becomes linear and $R_{xy}/H$ is almost $T-$independent, typical
for the ordinary Hall effect in the paramagnetic state of thin
films.

Cluster glass state is often accompanied by
superparamagnetic-type behavior. It
is described by the Langevin function\cite{superparamegnetism1}:

\begin{equation}\label{LangEq}
M \propto L=\coth\left(\frac{\mu
H}{k_{B}T}\right)-\frac{k_{B}T}{\mu H}.
\end{equation}
Here $\mu$ is the magnetic moment of an individual cluster.

\begin{figure}[t]
\includegraphics[width=18pc]{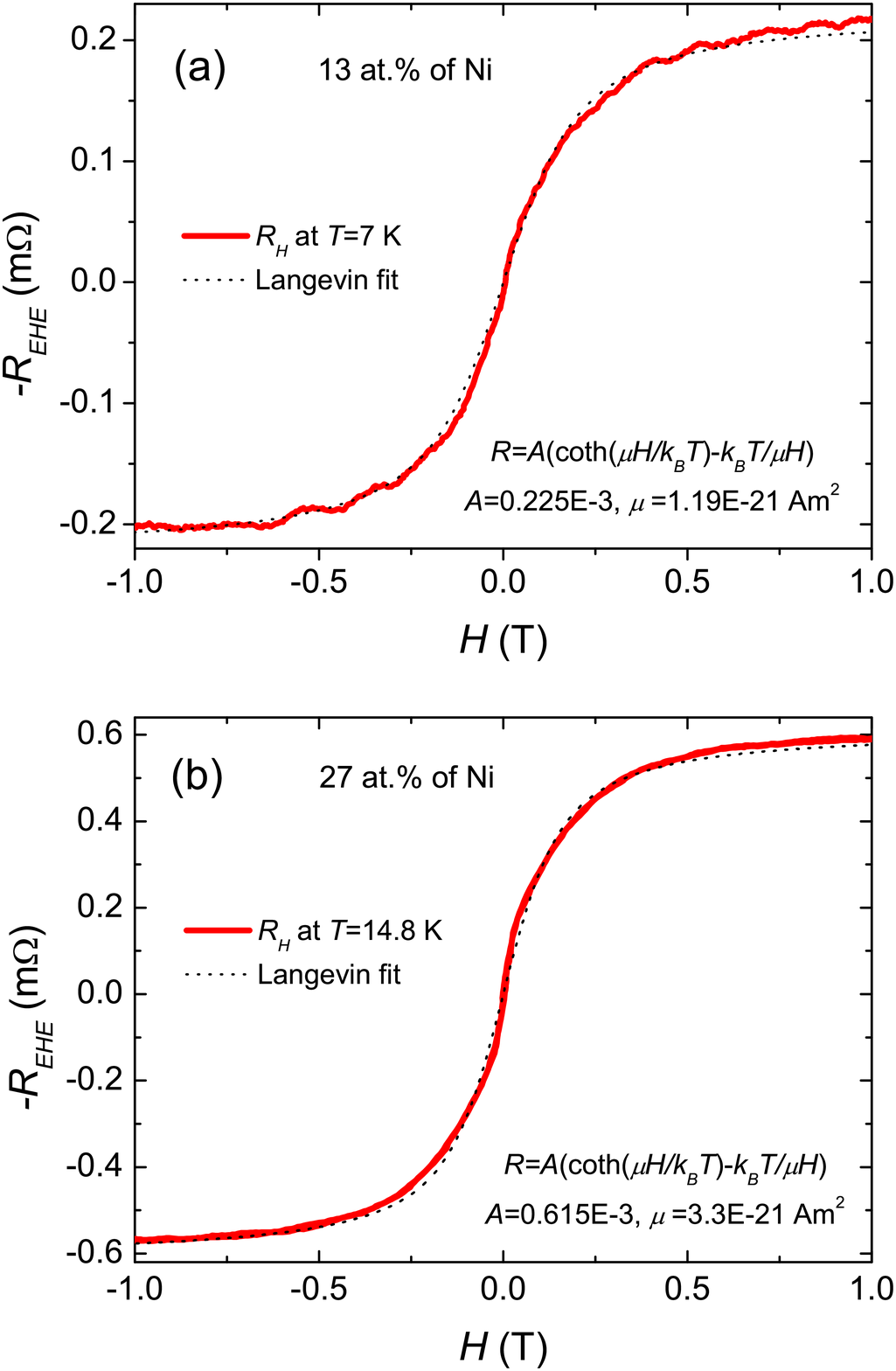}
\caption{\label{f12} (Color online) The extraordinary parts of the
Hall resistances $-R_{EHE}(H)$ for films with (a) 13 at.$\%$ of Ni
measured at $T$=7 K, and (b) 27 at.$\%$ of Ni measured at $T$=14.8
K. The dotted lines represent fits by the Langevin function.}
\end{figure}

The main panel of Fig. \ref{f11} shows $T$-evolution of
$-R_{xy}(H)$ for the film with 27 at.$\%$ Ni in the CG state at
$T>T_f$. It is seen that in the low field range from -0.3 T to 0.3
T, $R_{xy}(H)$ is strongly nonlinear and, therefore, has a
dominant EHE contribution, proportional to the magnetic moment of
individual magnetic clusters averaged over the volume of the
sample\cite{EHE_films}. To check the superparamagnetic scaling\cite{superparamegnetism}, in
the inset of Fig. \ref{f11} we plot $R_{xy}$ as a function of
$H/T$ in the low field range. It is seen that the scaling is
fairly good.

Figure \ref{f12} shows the EHE part of resistance, $R_{EHE}(H)$,
for films with (a) 13 at.$\%$ and (b) 27 at.$\%$ of Ni, at $T
\gtrsim T_f$. $R_{EHE}$ is obtained by subtracting a linear OHE
contribution, $R_{OHE}\propto H$, clearly seen at large $H$:
$R_{EHE}=R_{xy}-R_{OHE}$. Data are fitted with Eq. (\ref{LangEq}),
$R_{EHE}=AL$, using $\mu$ and $A$ as fitting parameters (dotted
lines). From the fits we obtain an average magnetization
$\mu=1.19\cdot10^{-21}$ A$\cdot$m$^{2}$ and $\mu=3.3\cdot
10^{-21}$ A$\cdot$m$^{2}$ for the films with 13 at.$\%$ and 27
at.$\%$ of Ni, respectively. Therefore, magnetic clusters are composed of about 128 and 356 $\mu_{B}$ for films with 13 and 27 at.$\%$ of Ni,
correspondingly. However, in contrast to fixed-size (structural) superparamagnetic particles, $\mu$
in NiPt films is not constant, but is gradually decreasing with
increasing temperature, which typical for the cluster glass state
with variable-size spin clusters.

\begin{figure}[t]
\includegraphics[width=18pc]{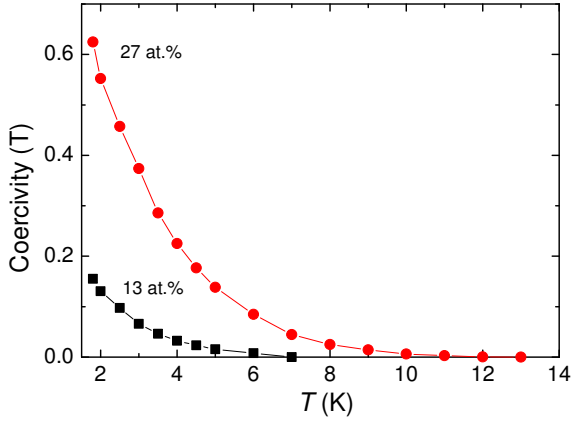}
\caption{\label{f20} (Color online) Coercivity, estimated from
$R_{xy}$ vs $H$ curves, at different temperatures for the films
with 13 (solid squares) and 27 (solid circles) at.$\%$ of Ni, with
magnetic field applied perpendicular to the film surface.}
\end{figure}

Figure \ref{f20} summarizes $T-$dependencies of the coercivity for
films with low Ni concentrations. The coercivity turns to zero at
the CG freezing temperature. This allows an accurate estimation of
$T_f\simeq $ 7 and 13 K for the films with 13 and 27 at.$\%$ Ni
respectively.

\subsection{Films with high Ni concentration}

\begin{figure}[t]
\includegraphics[width=18pc]{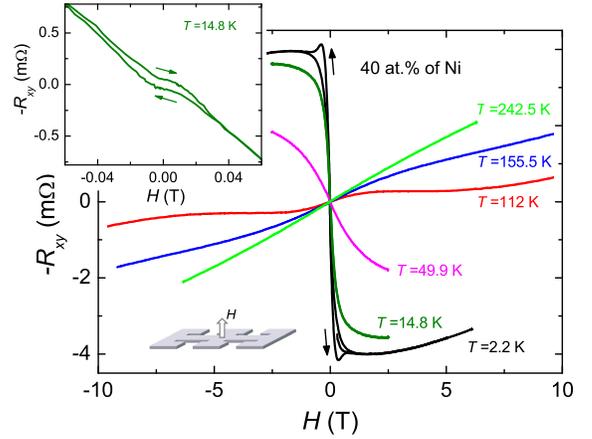}
\caption{\label{f13} (Color online) Magnetic field dependencies of
the Hall resistance for the film with 40 at.$\%$ of Ni measured at
different temperatures. The magnetic field is applied
perpendicular to the film. The inset shows $R_{xy}(H)$ at $T=14.8$
K in a small field range where inverse hysteresis is seen. Note
different signs of the EHE and the OHE at low $T$.}
\end{figure}

NiPt films with over-critical Ni concentration show distinctly
different behavior. In Fig. \ref{f13} we show $-R_{xy}(H)$ at
different $T$, for the film containing 40 at.$\%$ of Ni, which is
close to the critical concentration for the appearance of
ferromagnetism. At low $T$ the $-R_{xy}(H)$ curves decrease with
increasing field, opposite to that for sub-critical
concentrations, see Fig. 6. This means that the EHE coefficient
$R_1$ changes sign from electron-type (negative) at sub-critical,
to hole-type (positive) at the critical concentration. The EHE is dominant up to $H\simeq2$ T. At higher
fields the non-saturable OHE contribution becomes visible. The OHE coefficient $R_0$ is remaining electron-type, as in sub-critical films. Different signs of $R_0$ and $R_1$ lead to backbending of $R_{xy}(H)$ curves at high fields (see $R_{xy}(H)$
at $T=2.2$ K in Fig. \ref{f13}).

A new characteristic feature is seen at the lowest $T$=2.2 K: a
pronounced peak appears in $R_{xy}(H)$ at $H\simeq \pm$0.35 T. The
peak is hysteretic, i.e. depends on the field sweep direction. For
example, when the field is swept down from positive to negative
field, the peak appears at $H=-0.35$ T, but not at $H=+0.35$ T.
Such non-monotonic behavior of $R_{xy}$ is attributed to obliquely
canted magnetization\cite{Okamoto_oblique1,Okamoto_oblique2}. The
peak is smeared out already at $T$=3.8 K, and the hysteresis
vanishes at $T$=10.8 K. The hysteresis-free curves indicate an
in-plane easy axis of magnetization. At $T$=14.8 K, another
``flat" hysteresis in the reverse, clockwise direction, appears in
the low field region (see the inset in Fig. \ref{f13}). Such
behavior is similar to observations in
Ref.~\onlinecite{Okamoto_oblique1}, and is associated with
restoration of the canted magnetization, with the easy axis
flipping to the opposite side of the film, compared to the low-$T$
case. The hysteresis disappears at $T\simeq25$ K. At $T\geq 112$
K, the EHE coefficient changes sign to electron-type. At
$T\simeq240$ K the film becomes paramagnetic with linear
$R_{xy}(H)$.

\begin{figure}[t]
\includegraphics[width=18pc]{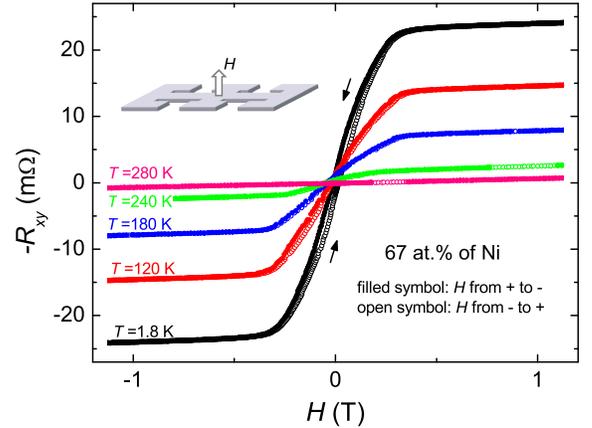}
\caption{\label{f12n} (Color online) Magnetic field dependencies of
the Hall resistance for the film with 67 at.$\%$ of Ni, at different temperatures. The magnetic field is applied
perpendicular to the film. The $R_{xy}(H)$ exhibit a clear
saturation up to $T_C \simeq 280$ K.}
\end{figure}

\begin{figure}[t]
\includegraphics[width=18pc]{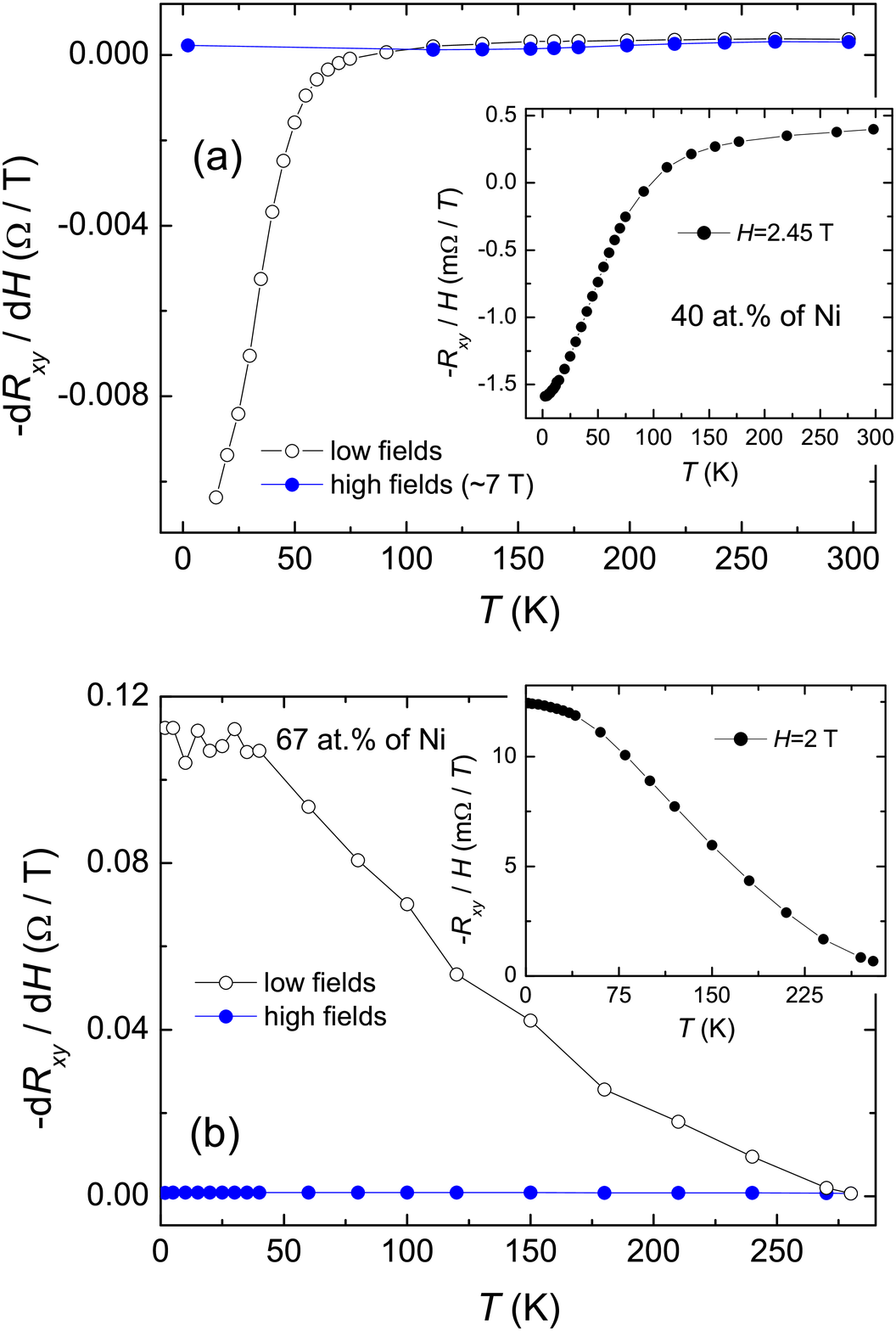}
\caption{\label{f13n} (Color online) Temperature dependencies of
$-dR_{xy}/dH$ at low (open) and high (closed symbols) fields,
perpendicular to films, for (a) 40 at.$\%$ and (b) 67 at.$\%$ of
Ni. The insets show normalized Hall resistances, at large fields.
Unlike the case of low Ni concentration, these are dominated by
EHE and exhibit strong $T$-dependence.}
\end{figure}

Figure \ref{f12n} shows $-R_{xy}(H)$ at different temperatures for
the film with the highest studied Ni concentration of 67 at.$\%$.
The curves show almost no hysteresis and a clear saturation. In
contrast to the film with 40 at.$\%$ of Ni, both the EHE and the
OHE has the same electron-type sign at all temperatures. The
saturation value of $R_{xy}$ at the lowest $T$ is seven times
larger than for the film with 40 at.$\%$ of Ni, shown in Fig.
\ref{f13}. The EHE contribution is dominant over the full range of
the applied magnetic field. This results in a relatively small $-dR_{xy}/dH$
at high fields [solid circles in Fig. \ref{f13n} (b)] and strong
temperature dependence of the saturation $R_{xy}$ [see the inset
of Fig. \ref{f13n} (b)]. The low field slope decreases with $T$,
and $R_{xy}(H)$ dependence becomes linear at the Curie temperature
$T_C \simeq 280$ K.

The main panels in Fig. \ref{f13n} show $T-$dependencies of $-dR_{xy}/dH$ for films with (a) 40 and (b) 67 at.$\%$ of Ni, at
small (open symbols) and large (closed symbols) fields. At low
temperatures, $R_{xy}(H)$ is strongly nonlinear, saturates at
relatively low fields, and is predominantly due to the EHE.
Consequently, $dR_{xy}/dH$ is large (positive for the film with 40
at.$\%$ of Ni and negative for the film with 67 at.$\%$ of Ni) at
low fields. The high-field slopes represent the OHE coefficient
$R_0$. It is small and negative in all cases. With increasing $T$,
the saturation magnetic field increases. As a result, the absolute
value of $dR_{xy}/dH$ decreases at low fields, but starts to
increase with increasing $T$ at high fields. The insets in Fig.
\ref{f13n} show values of $-R_{xy}/H$ at high fields. They have
strong temperature dependencies, as do low-field $R^{*}_{xy}/H$
shown in Fig. \ref{f06}. This is in contrast to films with low Ni
concentration, for which only low-field Hall effect had strong
$T-$dependence, while large field $R_{xy}$ was only weakly
$T-$dependent because it was dominated by the OHE. Clearly, for
NiPt thin films with high Ni concentrations, the contribution from
the EHE is substantial in the whole field range.

\subsection{Angular dependence of the anomalous Hall effect}

From Figs. \ref{f13} and \ref{f12n} it is seen that although films
with high Ni concentration are ferromagnetic and $R_{xy}(H)$ is dominated by the
extraordinary contribution $R_1 M$, the observed coercivity
remains very small. As mentioned in Sec. II C, such the behavior
is expected for films with the in-plane orientation of the easy
axis of magnetization.

\begin{figure}[t]
\includegraphics[width=18pc]{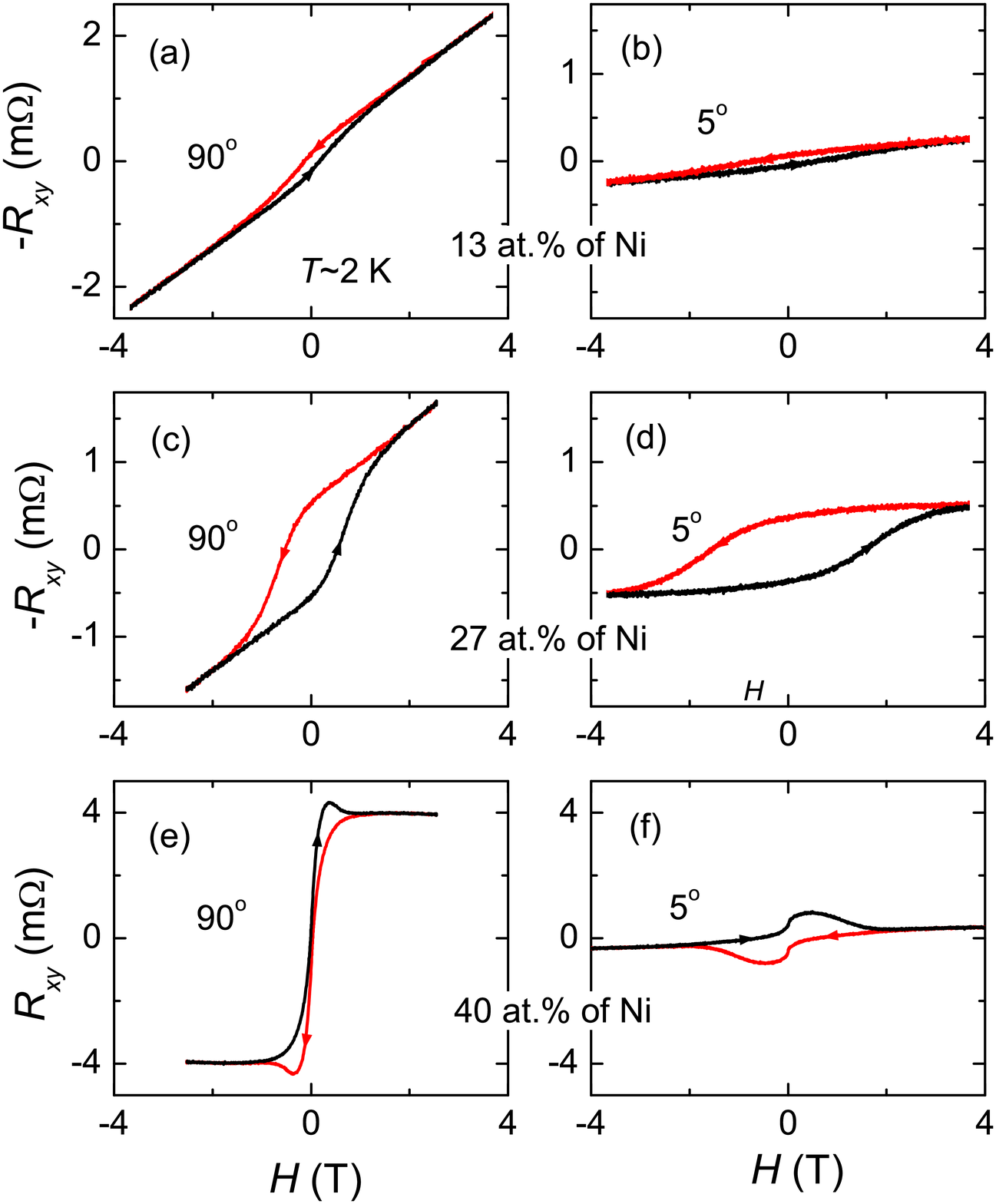}
\caption{\label{f14n} Hall resistances at $T\simeq 2$ K for
diluted NiPt films with (a,b) 13, (c,d) 27 and (e,f) 40 at.$\%$ of
Ni at angles 90$^{\circ}$ (field perpendicular to the film) and
5$^{\circ}$ between the film and the field. The $R_{xy}(H)$ curves
develop gradually with angle and collapse to $R_{xy}(H)\simeq 0$
at field parallel to the film. Note the non-monotonous $R_{xy}(H)$
of the ferromagnetic film with 40 at.$\%$ Ni, which is different
from the monotonous behavior of cluster-glass films with
sub-critical 13 and 27 at.$\%$ Ni concentration.}
\end{figure}

To better understand the orientation of magnetic anisotropy, we study angular dependence of the Hall effect. Figure \ref{f14n} shows $-R_{xy}(H)$ curves at
$T\sim$ 2 K for films with (a, b) 13, (c, d) 27, and (e, f) 40 at.$\%$ of Ni at angles (a, c, e) 90$^{\circ}$ (field
perpendicular to the film) and (b, d, f) 5$^{\circ}$ between the
field and the film surface. In all cases saturation values of
the Hall resistance decrease with decreasing angle because the
measured signal is proportional to the out-of-plain component of
the magnetic moment, which scales as the sine of angle.
Similarly, the saturation occurs at approximately the same
perpendicular component of the field. This leads to the seeming
stretching of the $R_{xy}(H)$ curves along the $H$-axis, inversely
proportional to the sine of the angle. The $-R_{xy}(H)$ curves in
diluted films, shown in Fig. \ref{f14n}, vary continuously with
the angle until they practically collapse to $R_{xy}(H)=0$ at
0$^{\circ}$, field parallel to the film.

\begin{figure}[t]
\includegraphics[width=18pc]{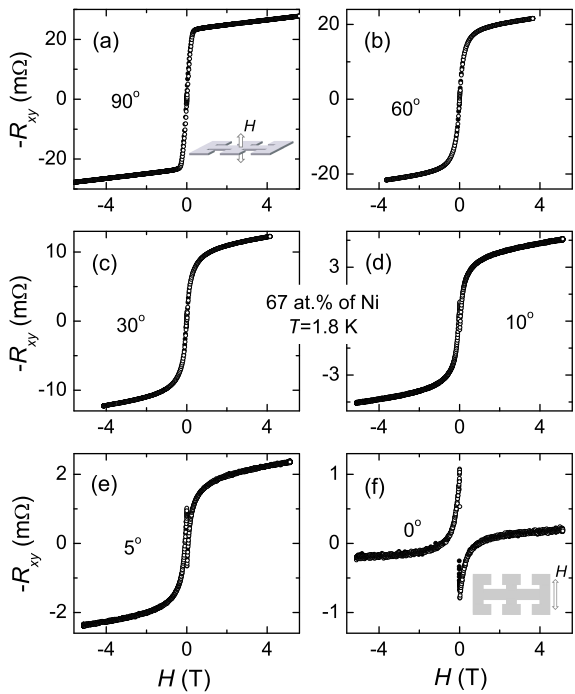}
\caption{\label{f17} Magnetic field dependencies of the Hall
resistances for the film with 67 at.$\%$ of Ni at $T$=1.8
K for different angles between the field and the film surface.
Switching of $R_{xy}$ at small angles and fields is associated
with the planar Hall effect.}
\end{figure}

Figure \ref{f17} shows $-R_{xy}(H)$ curves at $T$=1.8 K for the
film with 67 at.$\%$ of Ni at different angles between the field
and the film surface from (a) 90$^{\circ}$ (field perpendicular to
the film) to (f) $\sim 0^{\circ}$ (field parallel to the film).
Orientation of the field is sketched in panels (a) and (f). At the
angle of $10^{\circ}$ an abrupt switching of Hall resistance
appears in the low field region [see Fig. \ref{f17}(d)]. As the
angle decreases further, the values of
$R_{xy}$ between switchings slightly increase and reaches maximum at
$0^{\circ}$ [see Fig. \ref{f17}(f)]. For parallel field
orientation the saturation $R_{xy}$ at large fields becomes almost
zero. In this case, the magnetic moment is oriented in-plane and does not produce a measurable Hall voltage.
The switching of $R_{xy}$ at small fields is associated with the
planar Hall effect (PHE) \cite{Lu,Lu1}. The PHE appears when there
is a component of magnetization in the film plane. The
contribution from the PHE is zero when the magnetization is
parallel or perpendicular to the current \cite{Stinson}. That is
why it is zero at high fields when the sample is magnetized
perpendicular to the current flow direction [see Fig.
\ref{f17}(f)]. Note that a similar non-monotonous $R_{xy}(H)$ is
observed for the film with 40 at.$\%$ of Ni, but not observed for
more diluted films with 13 and 27 at.$\%$ of Ni, as seen from Fig.
\ref{f14n}. This again demonstrates different behavior of films
with low and high Ni concentrations.

\begin{figure}[t]
\includegraphics[width=18pc]{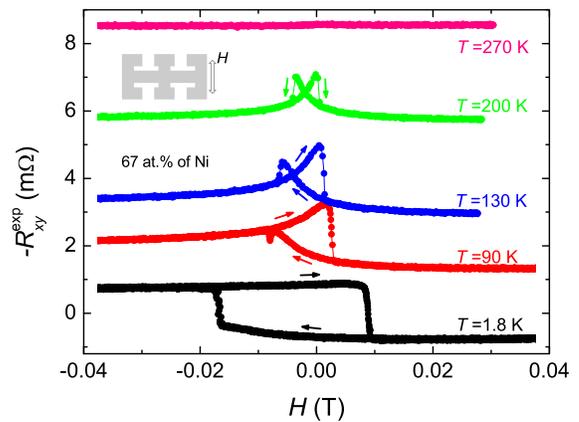}
\caption{\label{f18} (Color online) Measured Hall resistances in
small parallel magnetic fields at different temperatures for the
67 \% Ni film. The curves are offset for clarity. The presence of
both even- and odd-in-field contributions is seen, attributed to the planar and the extraordinary Hall effects, respectively.}
\end{figure}

Figure \ref{f18} represents a detailed view of the switching
process at different temperatures for the film with 67 at.$\%$ of Ni. It shows
$R^{\mathrm{exp}}_{xy}(H)$ of the same film for small in-plane
magnetic fields. The curves at different $T$ are offset for clarity. The measured
resistance contains both the even-in-field PHE and the
odd-in-field EHE contributions. The PHE is related to the
anisotropic magnetoresistance \cite{Stinson,Hurd}, which is even with
respect to the field direction. It is seen that the PHE
contribution is dominant at high $T$. However, at low $T$ there is
a clear odd-in-field EHE contribution. We remind that the measured
EHE signal is proportional to the out-of-plane moment. We
attribute EHE in nominally parallel fields to finite out-of-plane
stray fields appearing upon re-orientation of in-plane domains. As
temperature increases the $R^{\mathrm{exp}}_{xy}(H)$ becomes more
symmetrical indicating the decrease of the EHE contribution.

\subsection{The phase diagram}

In Table \ref{table1} we summarize magnetic properties of our
films: the Curie temperature $T_{C}$, the cluster-glass freezing
temperature $T_{f}$, cluster glass - to - paramagnet (PM)
transition temperature $T_{p}$, the saturated extraordinary Hall
resistivity
\begingroup\large$\rho$\endgroup$_{1}=R_{EHE}d$ measured at $T\sim 2$ K
and $H=2.5$ T, the longitudinal resistivity
\begingroup\large$\rho$\endgroup$_{xx}$ at $T\sim 2$ K and $H=0$ T
and the ordinary Hall coefficient $R_{0}$ at $T\sim 2$ K and 300
K, obtained from $dR_{xy}/dH$ at high fields. It is seen that
$T_{C}$, $T_{f}$, and $\rho_{1}$ increase rapidly (almost
exponentially) with increasing Ni concentration. However the
coercivity of films with high Ni concentration is much smaller
than that for films with low Ni concentration. This is caused by
reorientation of magnetic anisotropy from the out-of-plane to the
in-plane with increasing Ni concentration.

\begin{table*}
\caption{\label{table1} Parameters of studied NiPt thin films: the
Curie temperature $T_{C}$, the cluster glass freezing temperature
$T_{f}$, the cluster glass - to - paramagnetic transition
temperature $T_{p}$, the saturated extraordinary Hall resistivity
\begingroup\large$\rho$\endgroup$_{1}$ measured at $T\sim 2$ K
and $H=2.5$ T, the longitudinal resistivity
\begingroup\large$\rho$\endgroup$_{xx}$ at $T\sim 2$ K and $H=0$ T
and the ordinary Hall coefficient $R_{0}$ at $T\sim 2$ and 300 K.}
\begin{ruledtabular}
\begin{tabular}{cccccccc}
Ni
concentration&$T_{C}$&$T_{f}$&$T_{p}$&\begingroup\large$\rho$\endgroup$_{1}$
&\begingroup\large$\rho$\endgroup$_{xx}$&$R_{0}$($\sim$2
K)&$R_{0}$($\sim$300 K)
\\
(at.$\%$)&(K)&(K)&(K)&($10^{-3}\,\mu\Omega\,\mathrm{cm}$)
&($\mu\Omega\,\mathrm{cm}$)&($10^{-3}\,\mu\Omega\mathrm{\,cm\,T}^{-1}$)&($10^{-3}\,\mu\Omega\mathrm{\,cm\,T}^{-1}$)
\\
\hline
\vspace{-0.6pc}\\
13 &      &   7   & 60    &  -0.9   & 15.46  & -2.94 & -2.38\\
27 &      &   13  & 112   &  -2.7   & 27.31  & -2.16 & -2.26\\
40 & 3.8\footnotemark[1] & 24.8  & 242.5 &  19.45 & 35.56  & -1.14 & -1.55\\
67 & 280  &       &       &  -139.5 & 31.37  & -4.91 & -4.04\\
\end{tabular}
\end{ruledtabular}
\footnotetext[1]{Due to the coexistence of CG and FM at this concentration it is difficult to assess the true $T_{C}$ from the measured data.}
\end{table*}

\begin{figure}
\includegraphics[width=18pc]{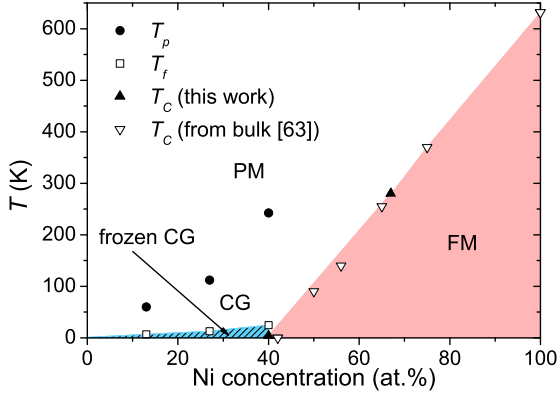}
\caption{\label{f21} (Color online) The obtained $T-x$ phase
diagram of Ni$_{x}$Pt$_{1-x}$ thin films. The films with low Ni
concentrations change the state from the paramagnetic (PM) at high
$T$, cluster-glass (CG) at intermediate $T$, and to frozen CG at
low $T$. The film with the critical Ni concentration of 40 at.$\%$
undergoes a double transitions from the PM to the CG and to the
ferromagnetic (FM) state with decreasing $T$. The film with 67
at.$\%$ of Ni shows a single phase transition from the PM to the
FM state at $T_C \sim$ 280 K. Open triangles represent $T_C$ of
bulk NiPt alloys from Ref.~\onlinecite{Tc}. The $T_C$ is vanishing
at the critical concentration $x_c \simeq 40\%$, indicating the
occurrence of the quantum phase transition at $T = 0$.}
\end{figure}

In Fig. \ref{f21} we present the magnetic phase diagram of NiPt
films. Ferromagnetism in the film with high 67 at.$\%$ Ni
concentration appears in a conventional manner through the
second-order PM-FM phase transition at the Curie temperature
$T_C$. At $T<T_C$ finite magnetization rapidly
starts to appear and there is no signature of the CG state above
$T_C$. The values of $T_{C}$ for the film with 67 at.$\%$ of Ni
are compared to $T_{C}$ of bulk NiPt alloys from
Ref.~\onlinecite{Tc}. Good agreement verifies the accuracy of
determination of Ni concentration using the EDS technique.

Diluted NiPt thin films demonstrate non-trivial properties. In
sub-critical films with 13 and 27 at.$\%$ of Ni with decreasing
temperature first a crossover from the PM to the CG state occurs
at $T_p$. In the CG state the saturation magnetization becomes
finite, however, the coercivity remains zero, as shown in Figs.
\ref{f11} and \ref{f12}. At $T_f$ the CG state freezes which leads
to appearence of finite coercivity and remanent magnetization at
the time scale of the experiment, see Fig. \ref{f07}.

Appearance of magnetism in the film with $40\%$ Ni requires one
more step. First a PM-to-CG crossover takes place at $T_p\simeq$
240 K. The CG freezes at $T_f\simeq $ 24.8 K, which leads to
appearance of a finite coercivity, as shown in the inset of Fig.
\ref{f13}. With further decrease of $T$ below $\sim$3.8 K the Hall
resistance starts to show a much larger hysteresis with a
non-monotonous peak, see the curve at $T=2.2$~K in Fig. \ref{f13}.
The angular dependence of the peak is resembling that for the
ferromagnetic film with 67 at.$\%$ of Ni, see Fig. \ref{f17} (f),
indicative for the in-plane anisotropy and the finite contribution
from the planar Hall effect due to appearance of ferromagnetic
domain walls. Therefore we ascribe the appearance of such
distinctly different behavior at low $T$ to transition of the film
in the ferromagnetic state at $T_C\simeq 3.8$~K. It is likely that
the observed sequential PM-CG-FM appearance of magnetism in this
films has a percolative mechanism \cite{Percolation}. In this case
spin clusters appear at $T_p \simeq$ 240 K, grow in size with
lowering temperature and finally form an infinite ferromagnetic
cluster at $T_C\simeq 3.8$ K, which is, however, coexisting with
the cluster glass state in the remaining finite-size clusters.

From Fig. \ref{f21} it is seen that the ferromagnetism disappears
$T_C\rightarrow 0$ at the critical concentration $x_c \simeq 40
\%$. Therefore, at this concentration the transition from the
normal (PM) to the ferromagnetic state occurs at $T=0$. This
is the quantum phase transition (QPT), which is driven by quantum,
rather than thermal fluctuations \cite{Varma,Vojta}. The QPT can
be viewed as a consequence of competition between qualitatively
distinct ground states \cite{Senthil}. The competition leads to
enhanced fluctuations around the critical point, which is
reflected in the maximum of the longitudinal resistivity
$\rho_{xx}$ for the 40\% Ni concentration, as seen from Table
\ref{table1}. However, from the diagram in Fig. \ref{f21} it is
seen that the QPT in NiPt films is not ideal: the observed
cluster-glass behavior in the sub-critical films indicates local
magnetic phase separation. Even though we can not exclude the
influence of random Ni disorder on formation of magnetic clusters
in our films, the commonality of such phase coexistence for
various types of magnetic QPT suggests that phase-separation may be generic
to QPT \cite{Vojta}. For example, the phenomenon is also
discussed in the context of the so-called pseudogap, coexisting
with high temperature superconductivity in cuprates
\cite{Senthil}.

As reported above, properties of NiPt thin films change
significantly at the critical point. Most remarkably, the
extraordinary Hall coefficient $R_{1}$ is changing sign. At low
and high Ni concentrations $R_{1}$ is electron-type which is
characteristic both for pure Ni and Pt. However, at the critical
concentration $R_1$ becomes hole-type (positive). The sign change of the EHE indicates that it
is not due to scattering, i.e., not extrinsic, but has an
intrinsic origin, related to the Berry phase and electron
structure of spin-polarized itinerant charge carriers
\cite{Nagaosa}. The presence of the hole-type contribution is also
reflected in the OHE. From Table \ref{table1} it is seen that the
value of $R_{0}$ has a minimum at the critical concentration $x_c
\simeq 40 \%$. Therefore, we suppose that the observed sign change
of the EHE at the critical Ni concentration is connected with the
quantum phase transition and is associated with reconstruction of
the electronic structure in the spin-polarization state of the
alloy.

\section{Conclusions}
In conclusions,  magnetic properties of sputtered NiPt thin films
with different Ni concentration were studied. Temperature,
magnetic field and angular dependencies of the Hall resistance
were analyzed. It was found that NiPt thin films with low,
sub-critical, Ni concentration show cluster-glass behavior at low
temperatures and exhibit perpendicular magnetic anisotropy below
the freezing temperature. Films with over-critical Ni
concentration are ferromagnetic with parallel anisotropy. At the
critical concentration the state of the film is strongly
frustrated: with decreasing temperature, the anisotropy rotates
from out-of-plane to in-plane and again out-of-plane and the
magnetism appears via consecutive percolation-type paramagnet -
cluster glass - ferromagnet transitions, rather than a single
second order phase transition. Most remarkably, the extraordinary
Hall effect changes sign at the critical concentration, while the
ordinary Hall effect does not. We suggest that this may be a
consequence of the quantum phase transition at $T=0$, associated
with reconstruction of the electronic structure in the
spin-polarization state of the alloy.

\section{Acknowledgments}
We are grateful to H. Frederiksen for help with thin film
deposition and to K. Jansson and O. Terasaki for assistance with
EDS characterization. Financial support from the Swedish Research
Council, the K.\&A. Wallenberg foundation and the SU-core facility
in nanotechnology is gratefully acknowledged.


\providecommand{\noopsort}[1]{}\providecommand{\singleletter}[1]{#1}%

\end{document}